\documentclass[aps,eqsecnum,epsf,preprint]{revtex4}

\usepackage[dvips]{graphicx}
\usepackage{amsmath,amssymb}
\usepackage{bm}
\newcommand{\BF}[1]{\mbox{\boldmath $#1$}}
\newcommand{\BFS}[1]{\mbox{\scriptsize\boldmath $#1$}}

\newcommand{\ve}[1]{{\bf #1}}
\begin{document}
\title{Induced Chern-Simons term of a paired electron state in the quantum Hall system}
\author{T. Aoyama}
\address{
Institute for Solid State Physics, University of Tokyo, 
5-1-5 Kashiwanoha, Kashiwa, Chiba 277-8581, Japan
}
\author{N. Maeda}
\address{
Department of Physics, Hokkaido University, 
Sapporo 060-0810, Japan
}

\date{\today}
\begin{abstract}
The induced Chern-Simons term for a paired electron state is calculated in the 
quantum Hall system by using a field theory on the von Neumann lattice. 
The coefficient of the Chern-Simons term, which is the Hall conductance, 
has not only the usual term proportional to a filling factor due to 
P (parity) $\&$ T (time reversal) symmetry breaking 
but also correction terms due to P $\&$ T $\&$ U(1) symmetry breaking. 
The correction term essentially comes from the Nambu-Goldstone 
mode and depends on an infrared limit. 
It is shown that the correction term is related to a topological number of a gap function 
in the momentum space. 



\end{abstract} 

\maketitle
\section{Introduction}

The half-filled states for each Landau level (LL) show amazing diversity. 
It is widely believed that the composite fermion state \cite{composite}, which 
has an isotropic Fermi surface, is realized at the half-filled lowest LL. 
At the half-filled second LL, 
an incompressible state with a small energy gap is observed. 
This is a very rare case of an even denominator fractional quantized Hall 
conductance \cite{pair_exp}. 
At the half-filled third and higher LLs, a highly anisotropic compressible state, 
which has a highly 
anisotropic longitudinal resistivity, is observed \cite{stripe_exp}. 
Theoretical studies suggest that this state is a striped state
\cite{stripe_the}. 
In the von Neumann lattice formalism, 
the striped state (striped Hall gas) 
is a compressible state which has anisotropic 
Fermi surfaces \cite{stripe_maeda}. 

In this paper, we concentrate attention on the half-filled second LL, in which 
the 5/2 fractional quantized Hall conductance is observed. 
Moore and Read proposed the pfaffian state which is represented by the Laughlin wave 
function with the pfaffian factor \cite{MooreRead,Read_Green}. 
The pfaffian state is considered as a spin polarized $p$ wave superconducting state of composite 
fermions \cite{GMB}. 
Experiments show that the $5/2$ quantized Hall state disappears by a
tilted magnetic field and anisotropic states are realized \cite{Pan_tile}.
Motivated by these theoretical proposals and experimental observations, 
we apply the BCS method  to a striped state with anisotropic Fermi surfaces. 
A mean field approach shows that a spin polarized $p$ wave like pairing 
state at a long-distance physics is more stable than a striped state
at the half-filled second LL \cite{maeda_pair}. 
Since a Nambu-Goldstone (NG) mode appears in the paired electron state and the Galilean 
invariance is spontaneously broken in the striped state, 
the quantization of the Hall conductance must be studied carefully. 

The Hall conductance is quantized topologically as a reflection of a fully gapped state. 
The integer quantized Hall conductance is represented by a topological 
number in a one-body problem in a magnetic field and a periodic potential \cite{TKNN}. 
In a field theoretical formalism \cite{ishikawa}, the Hall conductance can be 
represented by a topological number of an electron 
Green's function by using the Ward-Takahashi (WT) identity. 
The Hall conductance is the coefficient of the Chern-Simons term 
in the effective action \cite{jackiw}. 
It is proved that the Chern-Simons term is not renormalized perturbatively beyond 
the one-loop order if the vacuum has only massive modes \cite{CoHil,Abe}. 
In a superconducting state, however, the U(1) symmetry is spontaneously 
broken and the massless NG mode appears. 
Furthermore, the WT identity is anomalous due to the off-diagonal
condensation in particle-hole space. 
Therefore the long wavelength limit must be treated delicately and 
the quantization of the Hall conductance becomes nontrivial in the present case.  
For example, in a P (parity) $\&$ T (time reversal)
violating  superconductor, in which P $\&$ T are spontaneously 
broken without an external magnetic field, the Chern-Simons term is induced. 
The coefficient of this Chern-Simons term has infrared singularities 
due to the NG mode of the broken $U(1)$ symmetry and 
is related with a topological number \cite{Goryo}. 
Related examples are investigated in some field theoretical models \cite{wu,khare}. 

In this paper, 
we examine to calculate the effective action for a paired electron state 
(not a paired composite fermion state \cite{Read_Green}). 
At the electron one-loop order, only the Chern-Simons term is induced and 
the coefficient becomes $\frac{e^2}{2\pi}\nu$ at a filling factor $\nu$ after 
integrating out the NG mode. 
Beyond the one-loop order, not only the Chern-Simons term but also 
the Meissner term are induced. 
In order to include a vertex correction, the anomalous WT identity is applied 
in a gauge invariant manner \cite{Nambu}. 
The coefficient of the Chern-Simons term has corrections to $\frac{e^2}{2\pi}\nu$. 
The corrections include an infrared singularity and are related to a topological 
invariant of a gap function in the momentum space, intriguingly. 

The paper is organized as follows. 
In section \ref{BCS}, we review the BCS formalism in the von Neumann
lattice basis. 
In section \ref{conductance}, the low-energy effective action of a paired electron 
states is derived in the one-loop order. 
In section \ref{ward}, 
we investigate the effective action beyond the one-loop order with the help of the 
WT identity. 
Section \ref{summary} is a summary and discussion. 
In Appendix \ref{notation}, some useful relations including the Pauli's
matrices are provided. 
In Appendix \ref{LLmatrix}, 
the matrix elements between different LL indices in a current vertex are given. 
In Appendix \ref{example}, 
we present an example of an infrared limit in which the correction term 
appears in the Hall conductance. 

\section{BCS formalism}
\label{BCS}

In a strong perpendicular magnetic field $\ve{B}$, 
an electron guiding-center coordinate $(X,Y)$ of a cyclotron motion
with frequency $\omega_c=\frac{eB}{m}$ is non-commutative as
\begin{eqnarray}
[X,Y]&=&i \frac{a^2}{2\pi},
\label{eq:non}
\end{eqnarray}
where $a=\sqrt{\frac{2\pi\hbar}{eB}}$. 
The guiding-center coordinate is a conserved quantity in each LL energy 
$E_l=\hbar\omega_c(l+\frac{1}{2})$; 
thus $(X,Y)$ space generates the degeneracy within one LL. 
The non-commutativity introduces a cell structure of area $a^2$. 
The minimal uncertainty state in $(X,Y)$ space is given by the coherent state, 
namely the eigenstate of $X+iY$. 
The states become a minimal complete set when expectation values of $(X,Y)$ 
are on sites of a lattice whose unit cell area is $a^2$~\cite{von}.
We call this lattice the von Neumann lattice (vNL) \cite{ochiai}. 
A unit magnetic flux penetrates the unit cell of the vNL. 
The magnetic translational invariance 
is replaced by the lattice translational invariance, 
which provides a two-dimensional momentum $\ve{p}$ whose fundamental region is a
magnetic Brillouin zone (MBZ), $\vert p_i\vert\le\pi/a$. 
Moreover, 
a local $U(1)$ gauge symmetry in $\bf p$ space appears. 
Within a LL, a one-electron state is labeled by 
$\ve{p}$ and feels the $U(1)$ gauge field along its path in $\bf p$
space. 
More specifically,
the electron is on the base manifold of a two torus where the unit flux penetrates 
the surface of the torus. 
An electron field is expanded by 
the direct product of the LL state $|l\rangle$ 
and the momentum state $|\beta_{\ve{p}}\rangle$ as 
\begin{equation}
\Phi(\ve{r},t)=
\int_{\rm MBZ}\frac{d^2p}{(2\pi)^2}
\sum_{l=0}^\infty b_{l,\ve{p};t}\langle \ve{r}|l\otimes \beta_{\ve{p}}\rangle.
\end{equation}
The electron operator $b_{l,\ve{p}}$ obeys a twisted boundary condition 
$b_{l,\ve{p}+2\pi\ve{n}}=b_{l,\ve{p}}e^{i\phi(p,n)}$ with 
$\phi(p,n)=\pi(n_x+n_y)-n_yp_x$.
In the following, we take $\hbar=c=a=1$. 
We use the rectangular lattice $(X,Y)=(mr_s,n/r_s)$, where $m$, $n$ are 
integers and $r_s$ is an asymmetry parameter of the vNL. 
The notations $\hat{\bf k}=(r_s k_x,k_y/r_s)$ 
and $\tilde{\bf p}=(p_x/r_s,r_s p_y)$ are used. 

First of all, 
we review the BCS formalism in the vNL formalism \cite{maeda_pair}. 
In particular, we apply the Nambu representation of a spinless fermion,
namely, a spin polarized electron due to a strong magnetic field. 

\subsection{Hartree-Fock-Bogoliubov-de Gennes Hamiltonian and current vertex}

Let us consider a clean electron system in a planar space in a strong magnetic field. 
The electrons interact each other through the Coulomb interaction. 
The total Hamiltonian is given by $H=H_0+H_c$ where 
\begin{eqnarray}
H_0&=&\int d^2r \Phi^\dagger(\ve{r})\frac{(\ve{p}+e\ve{A})^2}{2m}\Phi(\ve{r}), \\
H_c&=&\frac{1}{2}\int d^2rd^2r^\prime 
\rho(\ve{r})V_c(\ve{r}-\ve{r}')\rho(\ve{r}^\prime).
\end{eqnarray}
Here $\rho(\ve{r})=\Phi^\dagger(\ve{r})\Phi(\ve{r})$, $V_c(\ve{r})=q^2/r$ 
and $q^2=\frac{e^2}{4\pi \epsilon}$ used as the energy unit. 
In the BCS formalism \cite{Schrieffer},
$H_c$ is decomposed into 
a mean field Hamiltonian $H_m$ and a residual Coulomb interaction 
$H_c-H_m$ as $H=H_0+H_m+(H_c-H_m)$. 
The Hartree-Fock-Bogoliubov-de Gennes (HFBd) Hamiltonian 
$H_{\rm HFBd}=H_0+H_m$ is defined by
\begin{equation}
H_{\rm HFBd}
=
\sum_{l=0}^{\infty}
\int_{\rm MBZ} \frac{d^2p}{(2\pi)^2} 
\Psi^\dagger_{l,\ve{p}}\BF{g}_l(\ve{p})\cdot\BF{\tau}\Psi_{l,\ve{p}}
\label{eq:bcs}
\end{equation}
with a spinor \cite{Nambu}
\begin{equation}
\Psi_{l,\ve{p}}=\frac{1}{\sqrt{2}}
\left(
\begin{array}{c}
b_{l,\ve{p}}\\
b^\dagger_{l,-\ve{p}}
\end{array}
\right).
\end{equation}
Here, $\tau_i$ is the Pauli's matrix, and $\BF{g}_l(\ve{p})
=({\rm Re}\Delta_l(\ve{p}),-{\rm
Im}\Delta_l(\ve{p}),\xi_l(\ve{p}))$ is a mean field vector in which 
$\Delta_l(\ve{p})$ is a superconductivity gap and $\xi_l(\ve{p})$ is an
energy spectrum of a normal state. 
The mean field vector will be determined by solving simultaneous
self-consistent equations later. 
The energy eigenvalues of $\BF{g}_l(\ve{p})\cdot\BF{\tau}$ are 
given by $\pm g_l(\ve{p})(\equiv\pm\vert
\BF{g}_l(\ve{p})\vert)$. 

A Green's function ${\cal G}_l(p)$ without a LL mixing is defined in a matrix form for the 
particle-hole space by
\begin{eqnarray}
-i\int dt_1 e^{ip_0(t_1-t_2)} \langle T\Psi_{l_1,\ve{p}_1;t_1}
\Psi^\dagger_{l_2,\ve{p}_2;t_2}
 \rangle
&=&
\delta_{l_1,l_2}{\cal G}_l(p)
\sum_n (2\pi)^2 \delta^2(\ve{p}_1-\ve{p}_2-2\pi\ve{n})e^{i\tau_3\phi(p,n)}
\nonumber\\
\end{eqnarray}
with 
\begin{equation}
{\cal G}_l(p)=\frac{p_0\tau_0+\BF{g}_l(\ve{p})\cdot\BF{\tau}}{p_0^2-g_l(\ve{p})^2+i\delta}
e^{i\tau_3 p_0\delta}.
\label{eq:bcspropa}
\end{equation}
Here, $\tau_0$ is a unit matrix and $\delta$ is a positive infinitesimal.
The boundary conditions for components of $\BF{g}_l(\ve{p})$ are non-trivial as
\begin{eqnarray}
\xi_l(\ve{p}+2\pi\ve{n})&=&\xi_l(\ve{p}),\\
\Delta_l(\ve{p}+2\pi\ve{n})&=&e^{2i\phi(p,n)}\Delta_l(\ve{p}),
\end{eqnarray}
according to the twisted periodicity of $b_{l,\ve{p}}$. 
This indicates that the gap function is a complex number. 

Next, a current operator is introduced in the matrix form for 
the particle-hole space. 
The Fourier transformed current operator 
$j^\mu(\ve{q})=e\int d^2r e^{-i\ve{q}\cdot\ve{r}} j^\mu(\ve{r})$ is represented by
\begin{equation}
j^\mu(\ve{q})
=
\int_{\rm MBZ}\frac{d^2p}{(2\pi)^2}\sum_{l_1,l_2} \Psi^\dagger_{l_1,\ve{p}} 
\Upsilon_{(0)l_1,l_2}^\mu(\ve{p},\ve{p}-\hat{\ve{q}})
\Psi_{l_2,\ve{p}-\hat{\ve{q}}},
\label{eq:bcscurr}
\end{equation}
with the current vertex in the matrix form 
\begin{equation}
\Upsilon_{(0) l_1,l_2}^\mu(\ve{p},\ve{p}-\hat{\ve{q}})
=
\left(
\begin{array}{cc}
f^\mu_{l_1,l_2}(q) & 0 \\
0 & -\{f^\mu_{l_1,l_2}(-q)\}^*
\end{array}
\right)
e^{-i \tau_3\int^p_{p-\hat{q}} \BFS{\alpha}({\bf k})\cdot d \ve{k}}
\label{eq:Namubuvertex}
\end{equation}
where the matrix element is given by 
\begin{eqnarray}
f^\mu_{l_1,l_2}(q)
&=&
\langle l_1 |\frac{1}{2}\{v^\mu,e^{i \BFS{q}\cdot\BFS{\chi}}\}| l_2
 \rangle.
\end{eqnarray} 
Here, $v^\mu=(1,-\omega_c \eta,\omega_c \xi)$ is an electron velocity 
and $\BF{\chi}=(\xi,\eta)$ is a relative coordinate operator. 
Clearly, $\{f^\mu_{l_1,l_2}(-q)\}^\ast=f^\mu_{l_2,l_1}(q)$ follows from 
the definition of $f^\mu_{l_1,l_2}(q)$ (see Appendix \ref{LLmatrix}). 

The current vertex is entirely different from the one in an 
electron system without a magnetic field. 
The spatial components of the current vertex are not 
represented by a momentum derivative of a free kinetic energy. 
Furthermore, 
the current operator has the non trivial phase factor 
$e^{-i \int^p_{p-\hat{q}} \BFS{\alpha}({\bf k})\cdot d \ve{k}}$ 
caused by the U(1) gauge field in $\bf p$ space. 
The phase factor is given in the Landau gauge 
$\BF{\alpha}(\ve{k})=(\frac{k_y}{2\pi},0)$ 
as 
\begin{eqnarray}
i \int_{\ve{p}}^{{\ve{q}}}{\BF{\alpha}(\ve{k})}\cdot d\ve{k}
=-\frac{i}{4\pi}(p-q)_x(p+q)_y.
\label{eq:mo_phase}
\end{eqnarray}
The field strength 
$\mathcal{B}=\nabla_{\ve{k}} \times \BF{\alpha}(\ve{k})=-\frac{1}{2\pi}$ 
is equivalent to the unit flux in the MBZ.

\subsection{Gap equation}

In the HFBd Hamiltonian, $\BF{g}_l({\bf p})$ is determined by 
solving simultaneous self-consistent equations of the electron self-energy part:
\begin{eqnarray}
\xi_l(\ve{p})
&=&
E_l+\int_{\rm MBZ}\frac{d^2q}{(2\pi)^2}
\frac{g(\ve{q})-\xi_l(\ve{q})}{2g(\ve{q})}V_{l}^{\rm HF}(q-p)-\mu, 
\label{eq:meaneq}
\\
\Delta_l(\ve{p})
&=&
\int_{-\infty}^{\infty}\frac{d^2q}{(2\pi)^2}
\frac{\Delta_l(-\ve{q})}{2g(\ve{q})}V_l(\tilde{q}-\tilde{p})
e^{
i \int_{\ve{p}}^{\ve{q}}2 \BFS{\alpha}(\ve{k})\cdot d\ve{k}
},
\label{eq:gapeq}
\end{eqnarray}
where  
$V_{l}^{\rm HF}(q)
=
\sum_{\ve{n}}
\left\{\tilde{V}_l\left(
\frac{\tilde{q}_y}{2\pi}+\tilde{n}_y,
\frac{\tilde{q}_x}{2\pi}+\tilde{n}_x
\right)\!
-
V_l(\tilde{q}-2\pi\tilde{n})
\!\right\}
$, 
$\tilde{V}_l(\ve{q})\equiv\int\frac{d^2q}{(2\pi)^2}V_l(\ve{p})
e^{i\ve{p}\cdot\ve{q}}$, 
$V(q)=\int d^2r e^{i\ve{q}\cdot\ve{r}}V_c(\ve{r})$
and $V_l(\ve{p})=\{f_{l,l}^0(p)\}^2V(p)$. 
Note 
that the gap equation (\ref{eq:gapeq}) has the gauge field $\BF{\alpha}({\bf k})$ twice as 
large as the one of 
the current vertex matrix Eq.~(\ref{eq:Namubuvertex}). 
By using Eq.~(\ref{eq:D}),
the gap equation is rewritten in a non-local form 
with $\bf p$ space covariant derivative 
$\ve{D}=-i{\nabla}_{\ve{p}}+\BF{\mathcal{A}(\ve{p})}$ as 
\begin{eqnarray}
\Delta_l(\ve{p})=-\int_{-\infty}^{\infty}
\frac{d^2q}{(2\pi)^2}
V_l(q)
e^{
i \hat{\ve{q}}\cdot\ve{D}
}
\frac{\Delta_l(\ve{p})}{2g(\ve{p})},
\label{eq:deltal}
\end{eqnarray}
where 
$\BF{\mathcal{A}}(\ve{p})=2\BF{\alpha}(\ve{p})$. 
This is because the gap function $\Delta(\ve{p})$ is a
condensation of the electron pair feeling the gauge field $2\BF{\alpha}(\ve{p})$. 
Thus, a LL state expansion associated with the gauge field $\BF{\mathcal{A}}(\ve{p})$ 
is useful to 
solve Eq.~(\ref{eq:deltal}). 
The $\bf p$ space LL state $\varphi_n$ is defined by
\begin{eqnarray}
D^2 \varphi_n(\ve{p})&=&e_n \varphi_n(\ve{p})
\end{eqnarray}
with an eigenvalue $e_n=\frac{2}{\pi}(n+\frac{1}{2})$. 
The eigenfunctions with the same boundary condition as 
$\Delta_l(\ve{p})$ are given by
\begin{eqnarray}
\varphi_{n}^{(2)}(\ve{p})
&=&
\frac{1}{\sqrt{n!}}\big( \frac{\pi}{2}\big)^{\frac{n}{2}}
(D_x-iD_y)^n \sqrt{2}e^{-\frac{p_y^2}{2\pi}}
\vartheta_2\left(\left.\frac{p_x+ip_y}{\pi}\right|2i\right),
\\
\varphi_{n}^{(3)}(\ve{p})
&=&
\frac{1}{\sqrt{n!}}\big( \frac{\pi}{2}\big)^{\frac{n}{2}}
(D_x-iD_y)^n \sqrt{2}e^{-\frac{p_y^2}{2\pi}}
\vartheta_3\left(\left.\frac{p_x+ip_y}{\pi}\right|2i\right).
\end{eqnarray}
These eigenfunctions have the following properties 
$\varphi_n^{(2,3)}(-\ve{p})=(-1)^n\varphi_n^{(2,3)}(\ve{p})$, 
$\varphi_n^{(2)}(p_x+\pi,p_y)=-\varphi_n^{(2)}(p_x,p_y)$, 
$\varphi_n^{(3)}(p_x+\pi,p_y)=\varphi_n^{(3)}(p_x,p_y)$, 
$\varphi_n^{(2)}(p_x,p_y+\pi)=e^{-ip_x}\varphi_n^{(3)}(p_x,p_y)$, 
and
$\varphi_n^{(3)}(p_x,p_y+\pi)=e^{-ip_x}\varphi_n^{(2)}(p_x,p_y)$.  
Since the gap has the odd parity, $\Delta(-\ve{p})=-\Delta(\ve{p})$, 
we expand $\Delta_l(\ve{p})$ and $\frac{\Delta(\ve{p})}{2g(\ve{p})}$ as 
\begin{eqnarray}
\Delta_l(\ve{p})
&=&
\sum_{n\geq1,i=2,3}c_n^{(i)}\varphi_{2n-1}^{(i)}(\ve{p}),
\\
\frac{\Delta_l(\ve{p})}{2g(\ve{p})}
&=&
\sum_{n\geq1,i=2,3}d_n^{(i)}\varphi_{2n-1}^{(i)}(\ve{p}).
\end{eqnarray}
The coefficients $c^{(i)}_n$ and  $d^{(j)}_n$ are determined self-consistently. 
Furthermore, 
the chemical potential $\mu$ is determined by a 
filling factor $\nu$ as
\begin{eqnarray}
\nu&=&\int_{-\infty}^{\infty}\frac{dp_0}{2\pi i}
\int_{\rm MBZ}\frac{d^2p}{(2\pi)^2}\frac{1}{2}{\rm Tr}\left\{\tau_3{\cal G}_{l_0}(p)
\right\}\\
&=&
l_0+\frac{1}{2}-\int_{\rm MBZ}\frac{d^2p}{(2\pi)^2}
\frac{\xi_{l_0}(\ve{p})}{2g(\ve{p})}, 
\nonumber
\end{eqnarray}
with the assumption that the condensation happens at 
the $l_{0}$ th LL. 
The trace is taken for both the LL and the particle-hole
space. 

We suppose that a normal state is a striped Hall gas 
with anisotropic Fermi surfaces in the Hartree-Fock approximation. 
Since the kinetic energy in a strong magnetic field is quenched into the LL energy, 
the electron states are infinitely degenerate and a Fermi surface does not 
exist in the absence of the interaction. 
In the striped Hall gas, the momentum dependent kinetic energy is 
spontaneously generated by the Coulomb interaction in the Hartree-Fock approximation. 
Then the degeneracy is lifted and anisotropic Fermi surfaces are formed in the MBZ. 
In an electron system without a magnetic field, on the other hand, 
a momentum dependent free kinetic term exists and 
a Fermi surface is formed according to the Pauli's exclusion principle. 
In such a well-known system, the Cooper's instability is triggered by a 
certain attractive interaction, and then a gap function is determined by 
a gap equation. 
In the present case, however, the kinetic energy and 
the pairing condensation are generated at the same time by the Coulomb interaction, as 
seen in Eqs.~(\ref{eq:meaneq}) and (\ref{eq:gapeq}). 

In the reference \cite{maeda_pair}, 
a self-consistent calculation was done by using an effective 
Hamiltonian and 
a solution at the half-filled $l=1$ LL was numerically obtained as 
\begin{eqnarray}
\xi_1(p_y)
&=&
-t_{\rm eff} \cos p_y -t_{(0,3)} \cos 3 p_y-\mu',
\label{eq:kai1}
\\
\Delta(\ve{p})
&=&
c^{(3)}_1\varphi_1^{(3)}(\ve{p})
+
c^{(3)}_3\varphi_3^{(3)}(\ve{p}),
\label{eq:kai2}
\end{eqnarray}
where $t_{\rm eff}=0.03$, $t_{(0,3)}=-0.0003$, 
$c_1^{(3)}=0.0104$, $c_3^{(3)}=-0.0018$, 
and $\mu'=-0.0006$.
The gap function behaves like $p$-wave at the long wavelength limit. 
In the following, calculations of the Hall conductance are 
carried out for a general mean field solution. 
Therefore, any solutions satisfying the self-consistent equations 
can be applied to following calculations.

We assume that the magnetic field is strong enough to neglect 
the LL mixing in the Green's function, and the normal state is the 
striped Hall gas at the $l_0$ th LL. 
Then the mean field vector is written as 
\begin{eqnarray}
\BF{g}_l(\ve{p})
&=&
\left\{
\begin{array}{c}
({\rm Re}\Delta(\ve{p}),-{\rm Im}\Delta(\ve{p}),\xi_{l_0}(\ve{p}))
\quad l=l_0,
\\
(0,0,E_l-\mu)
\quad l\neq l_0,
\end{array}
\right.
\label{eq:gvec}
\end{eqnarray}
where the Fermi energy lies at the $l_0$ th LL. 

\section{Effective action in One-loop order}
\label{conductance}

We study a long-distance behavior of the spontaneous $U(1)$ symmetry breaking state by 
calculating the current correlation function 
\begin{eqnarray}
\pi^{\mu\nu}(q)
&=&
-i (\mathcal{TS})^{-1}\int_{-\infty}^\infty
dt_1dt_2 e^{-iq_0(t_1-t_2)}
\langle
T j^\mu(\ve{q};t_1) j^\nu(-\ve{q};t_2)
\rangle
\end{eqnarray}
in the electron one-loop order, where 
$\mathcal{TS}$ is a product of total time and total area. 
Under a static homogeneous electric field $E_y$, 
the current density in $x$ direction is calculated by using the 
gauge field $a_0^{\rm ext}=y E_y$ as
$\langle J^x(x)\rangle
=i \int dx^\prime \pi^{x\nu}(x-x^\prime)a_\nu^{\rm ext}(x^\prime)
=-i\partial_y \pi^{x0}(q)|_{q=0} E_y$, 
where $\partial_y=\frac{\partial}{\partial q_y}$. 
Thus, the physical observable $\sigma_{xy}$ reflects the low-energy 
behavior of the slope of $\pi^{\mu\nu}(q)$ at the origin, 
which should be gauge invariant.

We consider an effective action of a fluctuating gauge field $a_\mu$
around an external gauge field $A_\mu$. 
The action of our system is microscopically written as
\begin{eqnarray}
S[\Phi,a]
=
\int d^3x \Phi^\dagger(x)
\left\{
i\partial_0-e a_0+\mu
-H_0(\ve{p}+e\ve{A}+e\BF{a})-H_c
\right\}\Phi(x).
\end{eqnarray}
This action is manifestly invariant under the $U(1)$ gauge transformation 
$\Phi(x)\rightarrow e^{ie\theta(x)}\Phi(x)$ and $a_\mu(x)
\rightarrow a_\mu(x)+\partial_\mu\theta(x)$. 
The mean field action in the HFBd approximation, which describes the spontaneous $U(1)$ symmetry breaking, 
is given by 
\begin{eqnarray}
S_{\rm HFBd}[\Psi,a]
&=&
S_{(0)}[\Psi]+S_{\rm int}[\Psi,a],
\\
S_{(0)}[\Psi]
&=&
\oint d^3p \Psi^\dagger(-p) \mathcal{G} ^{-1}(p)\Psi(p),
\\
S_{\rm int}[\Psi,a]
&=&
-\int d^3x
\{ej^\mu(x)a_\mu(x)+\frac{e\omega_c}{2}j^0(x)\BF{a}(x)\cdot\BF{a}(x)\}.
\label{eq:sint}
\end{eqnarray}
We use a short notation for the integral as 
$\oint d^3p\equiv
\int_{\rm R^1} dp_0
\int_{\rm T^2} d^2p
=
\int_{-\infty}^{\infty}\frac{dp_0}{(2\pi)}
\int_{\rm MBZ}\frac{d^2p}{(2\pi)^2}
$ and the contraction of sub-scripts and super-scripts for the 
space-time index is taken. 
Here, the electron field is a two-component spinor for the 
particle-hole space, and 
$\mathcal{G}^{-1}$ and $j_\mu$ are the inverse Green's function  
in Eq.~(\ref{eq:bcspropa}) and the current in Eq.~(\ref{eq:bcscurr}),
respectively. 
The second term in Eq.~(\ref{eq:sint}) is a diamagnetic term. 
The action $S_{\rm HFBd}[\Psi,a]$ does not hold the $U(1)$ gauge symmetry, 
because the pairing condensation breaks it. 
In order to recover the symmetry in the action \cite{Weinberg}, 
a NG mode $\theta$ is introduced by 
\begin{equation}
\Psi(x)\rightarrow e^{ie\tau_3\theta(x)}\Psi(x). 
\end{equation} 
Accordingly, 
the NG mode always couples with others  
by replacing $a_\mu$ with $a_\mu+\partial_\mu\theta$ in $S_{\rm HFBd}[\Psi,a]$. 
After integrating the electron field, 
the low-energy effective action takes the form as
\begin{equation}
S_{\rm eff}[a,\theta]
=
\int\frac{d^3q}{(2\pi)^3}
\left\{\frac{e^2}{2}
(a_{\mu,-q}-iq_\mu \theta_{-q})\pi^{\mu\nu}(q)(a_{\nu,q}+iq_\nu
\theta_{q})
-
\frac{e\omega_c \nu}{4\pi}
(\BF{a}_{-q}-i\ve{q} \theta_{-q})\cdot(\BF{a}_{q}+i\ve{q} \theta_{q})
\right\},
\label{eq:eff1}
\end{equation}
where the second term comes from the diamagnetic current part. 
The long-distance behavior is obtained by the Taylor expansion 
$\pi^{\mu\nu}(q)=\pi^{\mu\nu}(0)+q_\rho\partial^\rho\pi^{\mu\nu}(q)|_{q=0}
+\mathcal{O}(q^2)$ in the effective action. 
%%%%%%%%%%%%%%%%%%%%%%%%%%
\begin{figure}[hpt]
\includegraphics[width=5cm,clip]{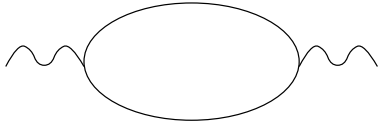}
\caption {
Feynman diagram for the current correlation function in one-loop order. 
The solid line stands for the Green function $\mathcal{G}$. 
The wavy line attached to solid line stands for the bare vertex. }
\label{fig:barebare}
\end{figure}
%%%%%%%%%%%%%%%%%%%%%%%%%%
The one-loop current correlation function $\pi^{\mu\nu}_{(0)}(q)$
is given by 
\begin{eqnarray}
\pi^{\mu\nu}_{(0)}(q)
&=&
ie^2
\oint d^3p\;\frac{1}{2}{\rm Tr}[\Upsilon_{(0)}^\mu(p,p-\hat{q})
\mathcal{G}(p-\hat{q})\Upsilon_{(0)}^\nu(p-\hat{q},p)\mathcal{G}(p)].
\label{eq:correlation2}
\end{eqnarray}
The corresponding 
Feynman diagram is shown in Fig.~\ref{fig:barebare}. 
The overall plus sign is due to the fermion one-loop. 

First, let us calculate $\pi^{\mu\nu}_{(0)}(0)$. 
The off-diagonal terms vanish immediately because 
$\pi^{0i}_{(0)}(0)$ contains 
${\rm Tr}^L[\Upsilon_{(0)}^0(p,p)\Upsilon_{(0)}^i(p,p)]=0$ 
and 
$\pi^{ij}_{(0)}(0)$ contains $\int dq
(\BF{g}_{l+1}\times\BF{g}_l)_z=0$ 
due to Eq.~(\ref{eq:gvec}). 
The diagonal terms create the gauge fields mass term and 
the kinetic term of the NG mode.
The direct calculation shows 
\begin{eqnarray}
 \pi^{00}_{(0)}(0)
&=&
\frac{e^2}{2}\int_{T^2} d^2 p\frac{|\Delta(\ve{p})|^2}{g^{3}_{l_0}}
\equiv v_n^2,
\\
 \pi^{ii}_{(0)}(0)
&=&
\omega_c
\left\{
\frac{e^2}{2\pi}\left(l_0+\frac{1}{2}\right)
-
\frac{e^2}{2\pi}\int_{T^2} d^2 p
\frac{\xi_{l_0}(\ve{p})}{2g_{l_0}(\ve{p})}
\right\}+\mathcal{O}\left(\frac{\vert\Delta\vert^2}{\omega_c}\right)
\nonumber\\
&=&
\frac{e^2\omega_c}{2\pi}\nu+\mathcal{O}\left(\frac{\vert\Delta\vert^2}{\omega_c}\right),
\end{eqnarray}
where we use Eq.~(\ref{eq:gvec}). 

Second, we calculate the $q$ linear terms
$\partial^\rho\pi^{\mu\nu}_{(0)}(q)|_{q=0}$. 
It is easy to show that 
$\partial^i \pi^{0j}_{(0)}(q)|_{q=0}=-\partial^i \pi^{j0}_{(0)}(q)|_{q=0}$, 
and
$\partial^0 \pi^{xy}_{(0)}(q)|_{q=0}=
- \partial^0 \pi^{yx}_{(0)}(q)|_{q=0}$
by 
$
\partial^\rho \Upsilon_{(0)}^{\mu}(p,p-\hat{q})|_{q=0}
=
-
\partial^\rho \Upsilon_{(0)}^{\mu}(p-\hat{q},p)|_{q=0}
$
and
${\rm Tr}\{\tau_3{\mathcal G}^2_{l+1}(p)\tau_0{\mathcal G}_l(p)\}
={\rm Tr}\{\tau_0{\mathcal G}^2_{l+1}(p)\tau_3{\mathcal G}_l(p)\}$. 
The calculation is straightforward and we obtain 
\begin{eqnarray}
 \partial^i \pi^{0j}_{(0)}(q)|_{q=0}
&=&
\frac{ie^2}{2\pi}\epsilon^{ij}
\left\{
l_0+\frac{1}{2}
-
\int_{T^2} d^2 p\frac{\xi_{l_0}(\ve{p})}{2g_{l_0}(\ve{p})}
-
\int_{T^2} d^2 p
\frac{E_{l_0}|\Delta(\ve{p})|^2}{2g^{3}_{l_0}(\ve{p})}
\right\}
+\mathcal{O}\left(\frac{\vert\Delta\vert^2}{\omega_c g_{l_0}}\right)
\nonumber
\\
&=&
i\left(\frac{e^2}{2\pi}\nu-C_{l_0} \right)\epsilon^{ij}
+\mathcal{O}\left(\frac{\vert\Delta\vert^2}{\omega_c g_{l_0}}\right),
\label{eq:dypi0x}
\\
\partial^0 \pi^{ij}_{(0)}(q)|_{q=0}
&=&
-\frac{ie^2}{2\pi}\epsilon^{ij}
\left\{
l_0+\frac{1}{2}
-
\int_{T^2} d^2 p\frac{\xi_{l_0}(\ve{p})}{2g_{l_0}(\ve{p})}
\right\}
+\mathcal{O}\left(\frac{\vert\Delta\vert^2}{\omega_c g_{l_0}}\right)
\nonumber
\\
&=&
-i\frac{e^2}{2\pi}\nu\epsilon^{ij}
+\mathcal{O}\left(\frac{\vert\Delta\vert^2}{\omega_c g_{l_0}}\right),
\label{eq:dypixy}
\end{eqnarray}
where $C_{l_0}=\frac{e^2}{2\pi}\int_{T^2} d^2 p
\frac{E_{l_0}|\Delta(\ve{p})|^2}{2g^{3}_{l_0}(\ve{p})}$. 
Apparently, 
$\partial^0 \pi^{ij}_{(0)}(q)|_{q=0}$ is different from 
$\partial^i \pi^{0j}_{(0)}(q)|_{q=0}$. 
The other terms, 
$\partial^\mu \pi^{\nu\nu}_{(0)}(q)|_{q=0}$, disappear in the effective action, 
since these terms are coefficients of parity odd terms $a_\nu(-q)q_\mu a_\nu(q)$. 
Next, we study whether the correction term $C_{l_0}$ survives or not after 
the NG field is integrated out. 

By substituting $\pi^{\mu\nu}_{(0)}(0)$ and 
$\partial^\rho\pi^{\mu\nu}_{(0)}(q)|_{q=0}$ into Eq.~(\ref{eq:eff1}), 
the effective action reads
\begin{eqnarray}
S_{\rm eff}[a,\theta]
&=&
\int\frac{d^3q}{(2\pi)^3}
\left\{
 \frac{v_n^2}{2} a_{0,-q}a_{0,q} 
+
\frac{ie^2\nu}{4\pi}\epsilon^{\mu\nu\rho}a_{\mu,-q}q_\nu a_{\rho,q}
-
i \frac{C_{l_0}}{2}\epsilon^{0ij}a_{0,-q}q_i a_{j,q}
+\frac{v_n^2}{2} q_0^2  \theta_{-q}\theta_{q}
\right.
\nonumber
\\
& &
+
i\left(v_n^2 q_0 a_{0,-q} 
+i C_{l_0} q_0\epsilon^{0ij}q_i a_{j,-q}
\right)\theta_{q}
\Big\}.
\end{eqnarray}
Here, we see that the $\pi^{ii}(0)$ cancels the diamagnetic current
term. 
After integrating out the NG field, 
we obtain the gauge invariant effective action
\begin{eqnarray}
S_{\rm eff}[a]
&=&
-i\int\frac{d^3q}{(2\pi)^3}
\frac{\nu e^2}{4\pi}\epsilon^{\mu\nu\rho}a_{\mu,q}q_\nu a_{\rho,-q}
\nonumber \\
&=&
\nu\frac{e^2}{2\pi}\int {d^3x} 
\frac{1}{2}\epsilon^{\mu\nu\rho}
a_{\mu}(x) \partial_\nu a_\rho(x).
\end{eqnarray}
As a result, 
at the fermion one-loop level, the induced low-energy effective action 
has merely the Chern-Simons term. 
Therefore, the Hall conductance in the one-loop order is given by
\begin{equation}
\sigma_{xy}^{\rm one-loop}=\frac{e^2}{2\pi}\nu.
\end{equation}
In a usual superconductivity, a gauge invariant mass term of 
gauge fields is induced. 
In the present case, however, it is not induced in the one-loop level. 
In the next section, higher order corrections to the current vertex part 
are investigated by using the WT identity. 

\section{Beyond one-loop calculation}
\label{ward}

In order to calculate the current correlation function beyond the one-loop order, 
we apply the WT identity to the current vertex part. 
To obtain the WT identity, 
it is quite useful to take 
the current basis in the vNL formalism following the reference \cite{ochiai}. 
In this basis, 
the electron annihilation operator is transformed by a 
unitary matrix as 
\begin{equation}
\tilde{b}_l({\bf p})=\sum_{l_1}U_{ll_1}({\bf p}) b_{l_1}({\bf p}). 
\end{equation}
The unitary matrix $U_{l_1l_2}({\bf p})$ is defined by 
\begin{eqnarray}
 U_{l_1l_2}({\bf p})
&=&
\langle l_1 |
e^{-i \tilde{p}_x \xi}e^{-i \tilde{p}_y \eta}
| l_2 \rangle.
\end{eqnarray}
The operator $U(\ve{p})$ is a translational operator for the 
electron relative coordinate as 
$U(\ve{p})f(\xi,\eta) U^\dagger(\ve{p})
=f(\xi+\frac{\tilde{p}_y}{2\pi},\eta-\frac{\tilde{p}_x}{2\pi})
$.
The phase factor with the U(1) gauge field $\BF{\alpha}(\bf p)$ 
in the density operator is eliminated 
by this transformation. 
The density operator is transformed into the form of one 
in the absence of the magnetic field, while the velocity is shifted in the current operator. 
In the current basis, the electron does not feel the U(1) gauge field $\BF{\alpha}(\bf p)$ 
but their cyclotron radii are changed. 
Actually, using the Hausdorff formula, 
the density and the current operators are rewritten as
\begin{eqnarray}
j^0(\ve{q})
&=&
\int_{T^2} d^2p\; \sum_{l_1,l_2}
b_{l_1,\ve{p}}^\dagger \{U^\dagger({\bf p})U({\bf p-\hat{q}})\}_{l_1,l_2}
b_{l_2,{\bf p-\hat{q}}}
\nonumber
\\
&=&
\int_{T^2}d^2 p\;
\sum_l \tilde{b}_{l,\ve{p}}^\dagger\tilde{b}_{l,\ve{p}-\hat{\ve{q}}},
\\
j^i(\ve{q})
&=&
\int_{T^2} d^2 p\; \sum_{l_1,l_2}
b_{l_1,\ve{p}}^\dagger\frac{1}{2}\{v^i,U^\dagger({\bf p})
U({\bf p-\hat{q}})\}_{l_1,l_2}
b_{l_2,{\bf p-\hat{q}}}
\nonumber
\\
&=&
\int_{T^2}d^2 p
\sum_{l_1,l_2}
\tilde{b}_{l_1,\ve{p}}^\dagger
\langle l_1 |
v^i+\frac{\omega_c}{2\pi}(\tilde{p}-\frac{q}{2})^i
| l_2 \rangle
\tilde{b}_{l_2,\ve{p}-\hat{\ve{q}}}.
\end{eqnarray}
In contrast to Eqs.~(\ref{eq:bcscurr}) and 
(\ref{eq:Namubuvertex}), the phase factor with the U(1) gauge field 
$\BF{\alpha}(\bf p)$ 
is absent and velocity operators are translated. 

In the Nambu representation, 
the electron field is transformed as
\begin{equation}
\tilde\Psi_l(\ve{p})=\sum_{l_1}{\cal U}_{ll_1}(\ve{p})\Psi_{l_1}(\ve{p}),
\end{equation}
with the unitary matrix 
\begin{equation}
{\cal U}({\bf p})=
\left(
\begin{array}{cc}
U({\bf p})&0\\
0& ^tU^*({\bf -p}) \\
\end{array}
\right).
\end{equation}
Then the density and current operator are written as
\begin{eqnarray}
j^0({\bf q}) &=& 
\int_{T^2} d^2 p \sum_{l_1,l_2}\Psi_{l_1}^\dagger({\bf p})
\{
{\cal U}^\dagger({\bf p}){\cal U}({\bf p-\hat{q}})
\}_{l_1,l_2}
\tau_3\Psi_{l_2}({\bf p-\hat{q}})
\nonumber
\\
&=&
\int_{T^2} d^2 p\sum_l\tilde\Psi^\dagger_l(\ve{p})
\tau_3\tilde\Psi_l(\ve{p}-\hat{\ve{q}}),
\\
j^i({\bf q}) &=&
\int_{T^2}d^2 p \sum_{l_1,l_2}
\Psi_{l_1}^\dagger({\bf p})\frac{1}{2}\{{\Upsilon}^i_{(0)},
{\cal U}^\dagger({\bf p})
{\cal U}({\bf p-\hat{q}})\}_{l_1,l_2}
\tau_3\Psi_{l_2}({\bf p-\hat{q}})
\nonumber
\\
&=&
\int_{T^2} d^2 p\sum_l\tilde\Psi^\dagger_l(\ve{p})
\langle l_1\vert{\Upsilon}_{(0)}^i \tau_3+
\frac{\omega_c}{2\pi}(p-\frac{q}{2})^i \tau_0\vert l_2\rangle
\tilde\Psi_l(\ve{p}-\hat{\ve{q}}),
\end{eqnarray}
where $\Upsilon^i_{(0)}=\Upsilon^i_{(0)}(p,p)$ which is given by
\begin{equation}
\Upsilon^i_{(0)}(p,p)=\left(
\begin{array}{cc}
v^i&0\\
0& ^t v^i \\
\end{array}
\right).
\end{equation}
The bare vertex function in this current basis is translated as
\begin{equation}
{\tilde\Upsilon}^\mu_{(0)}(p,p-q)=\tau_3\delta^{\mu0}+\left\{
{\Upsilon}^i_{(0)} \tau_3+
\frac{\omega_c}{2\pi}(p-\frac{q}{2})^i \tau_0\right\}
\delta^{\mu i}.
\end{equation}
In this way, 
the density vertex is diagonalized, hence, it is easy to obtain the WT identity.

\subsection{Current correlation function in the current basis}

\begin{figure}[hpt]
\includegraphics[width=5cm,clip]{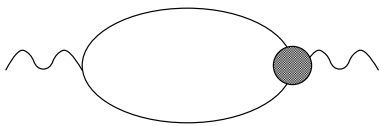}
\caption {Feynman diagram for the current correlation function of
 Eq.~(\ref{eq:corr0}).
The wavy line attached to dark circle stands for the full vertex. }
\label{fig:barefull}
\end{figure}

Before going to obtain and apply the WT identity, 
let us obtain a general formula of current 
correlation functions in the HFBd approximation. 
The current correlation functions are generally calculated as
\begin{eqnarray}
\pi^{\mu\nu}(q)
&=&
i e^2
\oint d^3p\; \frac{1}{2}{\rm Tr}
\{{\tilde\Upsilon}_{(0)}^\mu(p,p-\hat{q})
{\tilde{\mathcal G}}(p-\hat{q}){\tilde\Upsilon}^\nu(p-\hat{q},p)
{\tilde{\mathcal G}}(p)\}.
\label{eq:corr0}
\end{eqnarray}
where $\tilde{\mathcal G}(p)={\cal U}({\bf p}){\cal G}(p){\cal U}^\dagger({\bf p})$, 
and $\tilde{\Upsilon}^\nu$ is a full current vertex. 
The corresponding Feynman diagram is shown in Fig.~\ref{fig:barefull}. 
To avoid the double counting of diagrams, only one of two vertex parts is 
dressed. 

%%%%%%%%%%%%%%%%%%%%%%%%%%
%%%%%%%%%%%%%%%%%%%%%%%%%%

\begin{figure}[hpt]
\includegraphics[width=3cm,clip]{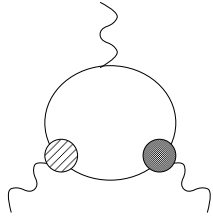}
\caption {Feynman diagram of Eq.~(\ref{eq:pisecond}). 
The wavy line attached to oblique circle stands for 
$\partial^\mu\mathcal{G}^{-1}$. }
\label{fig:dPibare}
\end{figure}

\begin{figure}[hpt]
\includegraphics[width=9cm,clip]{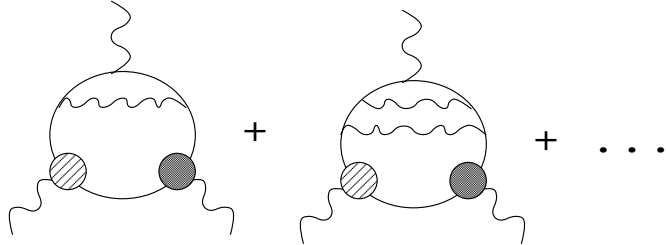}
\caption {Feynman diagram for the third term in Eq.~(\ref{eq:dPi}). 
Inner wavy line stands for the interaction potential $V(k)$. }
\label{fig:dPisum}
\end{figure}

Next we represent $\partial^\rho \pi^{\mu\nu}(q)\vert_{q=0}$ in a simple form. 
Differentiating Eq.~(\ref{eq:corr0}), we obtain
\begin{eqnarray}
\partial^\rho \pi^{\mu\nu}(q)\vert_{q=0}&=&
i e^2
\oint d^3p\; \frac{1}{2}{\rm Tr}
\{
\partial^\rho{\tilde\Upsilon}_{(0)}^\mu(p,p-\hat{q})\vert_{q=0}
{\tilde{\mathcal G}}(p){\tilde\Upsilon}^\nu(p,p)
{\tilde{\mathcal G}}(p)
\label{eq:dPi}\\
&&
+
{\tilde\Upsilon}_{(0)}^\mu(p,p)
\partial^\rho{\tilde{\mathcal G}}(p-\hat{q})\vert_{q=0}{\tilde\Upsilon}^\nu(p,p)
{\tilde{\mathcal G}}(p)
+
{\tilde\Upsilon}_{(0)}^\mu(p,p)
{\tilde{\mathcal G}}(p)\partial^\rho{\tilde\Upsilon}^\nu(p-\hat{q},p)\vert_{q=0}
{\tilde{\mathcal G}}(p)
\}.
\nonumber
\end{eqnarray}
The first term vanishes because it contains only double pole of $p_0$ 
owing to $\partial^\rho{\tilde\Upsilon}_{(0)}^\mu(p,p-\hat{q})\vert_{q=0}=(1-\delta^\mu_0)
\frac{\omega_c}{4\pi}\tau_0$. 
The second term becomes
\begin{equation}
i e^2
\oint d^3p\; \frac{1}{2}{\rm Tr}
\{
{\tilde\Upsilon}_{(0)}^\mu(p,p)
{\tilde{\mathcal G}}(p)
\partial^\rho{\tilde{\mathcal G}}^{-1}(p)
{\tilde{\mathcal G}}(p)
{\tilde\Upsilon}^\nu(p,p)
{\tilde{\mathcal G}}(p)\}
\label{eq:pisecond}
\end{equation}
by using the relation $-\partial^\rho\tilde{\mathcal{G}}=
\tilde{\mathcal{G}}\partial^\rho\tilde{\mathcal{G}}^{-1}\tilde{\mathcal{G}}$. 
This term is represented by the Feynman diagram in Fig.~\ref{fig:dPibare}. 
%%%%%%%%%%%%%%%%%%%%%%%%%%
%%%%%%%%%%%%%%%%%%%%%%%%%%
The third term in Eq.~(\ref{eq:dPi}) contains the derivative of the vertex part. 
In the HFBd approximation, we use the HFBd Green's function in Eq.~(\ref{eq:bcspropa}) 
in place of the full Green's function. 
Moreover, we use a dressed vertex defined by
\begin{eqnarray}
{\tilde\Upsilon}^\mu(p-\hat{q},p)&=&{\tilde\Upsilon}_{(0)}^\mu(p-\hat{q},p)
\nonumber\\
&&+
\int\frac{d^3q'}{(2\pi)^3}V(q')
\tau_3{\tilde{\cal G}}(p-\hat{q}-\hat{q}')
{\tilde \Upsilon}^\mu(p-\hat{q}-\hat{q}',p-\hat{q}'){\tilde{\cal G}}(p-\hat{q}')
\tau_3,
\label{eq:fullvertex}
\end{eqnarray}
in place of the full current vertex. 
Then the derivative of the vertex part is given by
\begin{eqnarray}
\partial^\rho{\tilde\Upsilon}^\mu(p-\hat{q},p)\vert_{q=0}&=&
\partial^\rho{\tilde\Upsilon}_{(0)}^\mu(p-\hat{q},p)\vert_{q=0}
\nonumber 
\\
&&+
\int\frac{d^3q'}{(2\pi)^3}\{V(q')
\tau_3\partial^\rho{\tilde{\cal G}}(p-\hat{q}-\hat{q}')\vert_{q=0}
{\tilde \Upsilon}^\mu(p-\hat{q}',p-\hat{q}'){\tilde{\cal G}}(p-\hat{q}')
\tau_3
\nonumber\\
&&+V(q')
\tau_3{\tilde{\cal G}}(p-\hat{q}')
\partial^\rho{\tilde \Upsilon}^\mu(p-\hat{q}-\hat{q}',p-\hat{q}')\vert_{q=0}
{\tilde{\cal G}}(p-\hat{q}')
\tau_3
\}.
\label{eq:dPi2}
\end{eqnarray}
By substituting Eq.~(\ref{eq:dPi2}) into the third term of Eq.~(\ref{eq:dPi}), 
the first term of them vanishes because of the same reason of 
the disappearance of the 
first term in Eq.~(\ref{eq:dPi}). 
We again substitute Eq.~(\ref{eq:dPi2}) into the remaining terms, and 
repeat the same procedure. 
As a result, 
we obtain the infinite series of Feynman diagrams as shown 
in Fig.~\ref{fig:dPisum}. 
%%%%%%%%%%%%%%%%%%%%%%%%%%
%%%%%%%%%%%%%%%%%%%%%%%%%%
In this way, 
combining the infinite series with Eq.~(\ref{eq:dPi2}), 
the bare vertex part is 
renormalized to the dressed vertex part of Eq.~(\ref{eq:fullvertex}). 

Consequently, we obtain the following useful formula 
\begin{eqnarray}
\partial^\mu\pi^{\nu\rho}(q)\vert_{q=0}
=
ie^2 \oint d^3p\; \frac{1}{2}{\rm Tr}
\{\partial^\mu{\tilde{\mathcal G}}^{-1}(p){\tilde{\mathcal G}}(p)
{\tilde\Upsilon}^\nu(p,p) 
{\tilde{\mathcal G}}(p){\tilde\Upsilon}^\rho(p,p){\tilde{\mathcal G}}(p)\}, 
\label{eq:dPifull}
\end{eqnarray}
%%%%%%%%%%%%%%%%%%%%%%%%%%
\begin{figure}[hpt]
\includegraphics[width=3cm,clip]{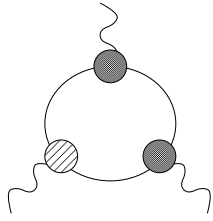}
\caption {Feynman diagram for $\partial^\mu\pi^{\nu\rho}$.} 
\label{fig:dPifull}
\end{figure}
%%%%%%%%%%%%%%%%%%%%%%%%%%
whose Feynman diagram is shown in Fig.~\ref{fig:dPifull}. 
Compared with Eq.~(\ref{eq:pisecond}), 
the bare vertex is replaced by the dressed vertex part. 
In the following subsection we calculate this term by using the WT identity. 

\subsection{Ward-Takahashi identity}

In the current basis, the equal-time commutation relation of the 
density operator 
and the electron operator becomes a simple form 
\begin{eqnarray}
 [j^0(\ve{q}),{\tilde b}_l(\ve{p})]=-{\tilde b}_l(\ve{p}-\hat{\ve{q}}),
\end{eqnarray}
as an ordinary electron case. 
Using this relation and the current conserving law $q_\mu j^\mu(q)=0$, 
we obtain the WT identity in the current basis as
\begin{eqnarray}
q_\mu\tilde\Upsilon^\mu(p,p-\hat{q})
&=&
\tilde{\mathcal G}^{-1}(p)\tau_3-\tau_3\tilde{\mathcal
G}^{-1}(p-\hat{q})
.
\end{eqnarray}
In a zero momentum limit, the off-diagonal part of $\tilde{\mathcal
G}^{-1}$ gives infrared singularities to the vertex function as seen in 
$
q_\mu\tilde\Upsilon^\mu(p,p-\hat{q})
=
\{
\tilde{\mathcal G}^{-1}(p)- \tilde{\mathcal G}^{-1}(p-\hat{q})
\}\tau_3
+
\tilde{\mathcal G}^{-1}(p-\hat{q})\tau_3
-\tau_3 \tilde{\mathcal G}^{-1}(p-\hat{q})
$. 
As a prescription to use the derivative form of the WT identity, 
we separate the regular part and the singular part in $\tilde{\Upsilon}^\mu$ as 
\begin{equation}
\tilde\Upsilon^\mu=\tilde\Upsilon^\mu_{\rm regular}+\tilde\Upsilon^\mu_{\rm singular},
\end{equation}
in which the regular part satisfies
\begin{eqnarray}
{\tilde\Upsilon}_{\rm regular}^\mu(p,p)
&=&
\partial^\mu{\tilde{\mathcal G}_{\rm reg}}^{-1}(p)\tau_3.
\label{eq:wt_r}
\end{eqnarray}
Here 
${\tilde{\mathcal G}_{\rm reg}}(p)$ is a diagonal part of the Green's function in 
the particle-hole space. 
The singular part originates from the NG mode propagation in an
internal line for the current vertex correction. 
This contribution is recovered by coupling of the NG mode with the gauge field. 
Hence, 
we use Eq.~(\ref{eq:wt_r}) in 
place of the full vertex part in the current correlation function. 

\subsection{$\pi^{\mu\nu}(0)$ and $\partial^\mu \pi^{\nu\rho}(q)|_{q=0}$}

First, let us calculate $\pi^{\mu\nu}(0)$. 
With the help of the WT identity Eq.~(\ref{eq:wt_r}), 
the current correlation function at $q=0$ is given by 
\begin{equation}
\pi^{\mu\nu}(0)=
ie^2\oint d^3p\;\frac{1}{2}{\rm Tr}\{{\tilde\Upsilon}_{(0)}^\mu(p,p)
{\tilde{\mathcal G}}(p)
\partial^\nu {\tilde{\mathcal G}}_{\rm reg}^{-1}(p)\tau_3{\tilde{\mathcal G}}(p)\}.
\end{equation}
By using the relation in Appendix~\ref{notation}, 
we can show that $\pi^{\mu\nu}(0)=0$ for $\mu\neq\nu$ and 
\begin{equation}
\pi^{00}(0)={e^2}
\int_{T^2}d^2 p\; \frac{\vert\Delta({\bf p})\vert^2}{2\{g({\bf p})\}^3}
\equiv v_n^2.
\end{equation}
In this section we omit the LL index $l_0$ in $\BF{g}_{l_0}$ for
simplicity. 
$\pi^{ii}(0)$ is given by
\begin{equation}
\pi^{ii}(0)=\omega_c\frac{e^2}{2\pi}\nu-\frac{e^2}{2\pi}
\int_{T^2}d^2 p\; \frac{(2l_0+1)\vert\Delta({\bf p})\vert^2}{2g({\bf p})}
\end{equation}
up to $\mathcal{O}(\vert\Delta\vert^2/\omega_c)$. 
The first term of $\pi^{ii}(0)$ 
is canceled with the diamagnetic term and the second term gives 
the mass term of gauge fields.

Next, we calculate $\partial^\mu \pi^{\nu\rho}(q)|_{q=0}$. 
Using Eq.~({\ref{eq:dPifull}}) and the WT identity, one finds 
\begin{eqnarray}
\partial^\mu \pi^{\nu\rho}(q)|_{q=0}=ie^2\oint d^3p\;\frac{1}{2}{\rm Tr}
\{\partial^\mu{\tilde{\mathcal G}}^{-1}(p){\tilde{\mathcal G}}(p)
\partial^\nu {\tilde{\mathcal G}}_{\rm reg}^{-1}(p)
\tau_3
{\tilde{\mathcal G}}(p)\partial^\rho{\tilde{\mathcal G}}_{\rm reg}^{-1}(p)
\tau_3
{\tilde{\mathcal G}}(p)\}. 
\end{eqnarray}
This is similar to the topological invariant obtained for the 
integer quantum Hall state \cite{ishikawa,ochiai}. 
In the integer quantum Hall state, the Hall conductance is topologically quantized and 
not renormalized beyond the electron one-loop order perturbatively. 
In the present case, on the other hand, the above term is not topological 
invariant due to the broken U(1) symmetry. 
In subsection E, we will extract the topological invariant from this term 
in the limit $\vert\Delta(\ve{p})\vert\rightarrow 0$. 

By tedious calculations with the help of the relation 
in Appendix~\ref{notation} and \ref{LLmatrix},
we can show that 
$\partial^i\pi^{ij}=0$, 
and
\begin{equation}
\partial^0\pi^{ij}
=
i e^2 \epsilon^{ij}\oint d^3p\; \frac{1}{2}{\rm Tr}
\{\frac{i}{2\pi}\tau_3{\mathcal G}(p)
+
\partial^x{\mathcal G}^{-1}_{\rm reg}\tau_3{\mathcal G}
\partial^y{\mathcal G}_{\rm reg}^{-1}\tau_3
{\mathcal G}^2\},
\end{equation}
up to $\mathcal{O}(\vert\Delta\vert^2/\omega_c g)$.  
The first term is proportional to a filling factor and the second term vanishes. 
We also obtain
\begin{eqnarray}
\partial^i\pi^{j0}
&=&i e^2 \oint\!\! d^3p \; \frac{1}{2}{\rm Tr}
\left\{\frac{i}{2\pi}\epsilon_{ij}\tau_3{\cal G}(p)
+
(\partial^i{\cal G}^{-1}+\mathcal{A}^i(\ve{p})\tilde\Delta)
{\cal G}\partial^j{\cal G}_{\rm reg}^{-1}\tau_3{\cal G}\tau_3{\cal G}
\right\}
\nonumber\\
&&
-i\epsilon_{ij}\frac{e^2}{2\pi}\int_{T^2}\!\!d^2p\; \frac{E_{l_0}\vert
\Delta\vert^2}{2\{g(\ve{p})\}^3}
\label{eq:ij0}
\end{eqnarray}
up to $\mathcal{O}(\vert\Delta\vert^2/\omega_c g)$, 
where $\tilde\Delta=g_x\tau_2-g_y\tau_1$. 
The first term is proportional to a filling factor. 
The second term is expressed by 
\begin{equation}
ie^2\int_{T^2}d^2p\;\frac{\{g_{x}(\ve{p})\partial^i 
g_{y}(\ve{p})-g_{y}(\ve{p})\partial^i g_{x}(\ve{p})
-\mathcal{A}^i(\ve{p})\vert\Delta\vert^2\}
\partial^j 
g_{z}(\ve{p})}{4\{g(\ve{p})\}^3}
\equiv
-i\alpha^{ij}
\label{eq:alphaij}
\end{equation}
which is related to a topological number in the momentum space. 
The term proportional to ${\bf \mathcal{A}}^i(\ve{p})$ 
in the numerator guarantees the
periodicity of the integrand. 
The third term in Eq.~(\ref{eq:ij0}) is proportional to $\omega_c$. 

\subsection{Effective action beyond one-loop}

Following section \ref{conductance}, 
the effective action of gauge fields $a_\mu$ and the NG
mode $\theta$ is given in a gauge invariant form as 
\begin{eqnarray}
S_{\rm eff}[a,\theta]&=&
\int d^3x
\left\{ \nu\frac{e^2}{2\pi}\frac{1}{2}\epsilon^{\mu\nu\rho}
a_\mu\partial_\nu a_\rho
+
(-\alpha^{ji}+C_{l_0}\epsilon^{ij})(a_i+\partial_i\theta)\partial_j 
(a_0+\partial_0\theta)\right.
\nonumber\\
&&\left.+\frac{v_n^2}{2}(a_0+\partial_0\theta)^2
-\frac{(c_s v_n)^2}{2}(a_i+\partial_i\theta)^2
\right\}.
\label{eq:Seffath}
\end{eqnarray}
Here, 
\begin{eqnarray}
(c_sv_n)^2
&=&
\frac{e^2}{2\pi}\int_{T^2}d^2p\frac{(2l_0+1)\vert\Delta\vert^2}{2g}.
\end{eqnarray}
After integrating the NG mode, we obtain the gauge invariant 
effective action as 
\begin{eqnarray}
S_{\rm eff}[a]&=&
\int d^3x
\left\{
\left(\nu\frac{e^2}{2\pi}
-
\frac{c_s^2(-\epsilon^{ij}\alpha^{ki}\partial_k\partial_j+C_{l_0}\partial_i^2)}
{\partial_0^2-c_s^2\partial_i^2}
\right)
\frac{1}{2}\epsilon^{\mu\nu\rho}a_\mu
\partial_\nu a_\rho\right.
\label{eq:eff_action}
\\
&&+\left.
\frac{v_n^2}{2}
\left(
(a_0)^2-c_s^2(a_i)^2+\frac{(\partial_0 a_0-c_s^2\partial_i a_i)^2}
{\partial_0^2-c_s^2\partial_i^2}
\right)\right \}.
\nonumber
\end{eqnarray}
The first term is the induced Chern-Simons term which is gauge invariant. 
The second term can be written by a quadratic term of transverse
components of gauge fields 
$a_0^{\rm T}\equiv a_0-\partial_0(\partial_0 a_0-c_s^2\partial_i
a_i)/(\partial_0^2-c_s^2\partial_i^2)$ and  
$a_i^{\rm T}\equiv a_i-\partial_i(\partial_0 a_0-c_s^2\partial_j
a_j)/(\partial_0^2-c_s^2\partial_i^2)
$
as $\frac{v_n^2}{2}\{(a_0^{\rm T})^2-c_s^2 (a_i^{\rm T})^2\}$. 
Since the transverse gauge field is invariant under the $U(1)$ gauge 
transformation, 
the obtained action has the manifest $U(1)$ gauge symmetry. 
The infrared singularity in the induced Chern-Simons term corresponds to 
the contribution of the NG mode which is involved in 
$\tilde{\Upsilon}^\mu_{\rm singular}$. 

To obtain the low-energy physics, we must take $\partial^\mu\rightarrow0$ 
limit in the coefficient of the induced Chern-Simons term. 
The order of zero momentum limit is delicate, because 
$\lim_{\partial_0\rightarrow 0}$ and
$\lim_{\partial_i\rightarrow 0}$ do not commute. 
In the limit $\partial_i\rightarrow0$ before taking $\partial_0
\rightarrow 0$, 
the effective action becomes
\begin{equation}
\lim_{\partial_0\rightarrow 0}
\lim_{\partial_i\rightarrow 0}
S_{\rm eff}[a]
=
\int d^3x\left[\nu\frac{e^2}{2\pi}\frac{1}{2}
\epsilon^{\mu\nu\rho}a_\mu\partial_\nu a_\rho
+\frac{v_n^2}{2}\{(a^{\rm T}_0)^2-c_s^2(a_i^{\rm T})^2\}
\right].
\end{equation}
Thus the Hall conductivity is not renormalized in this limit. 
In contrast to the one-loop order, the Meissner term appears. 
This infrared limit corresponds to the experimental situation of 
the quantum Hall effect, because a spatial homogeneous external electric field is added to the
system adiabatically. 
On the other hand, for the inverse order, 
the correction $C_{l_0}$ and $\alpha^{ij}$ appear in the Hall
conductance as  
\begin{eqnarray}
\lim_{\partial_x\rightarrow 0}
\lim_{\partial_y\rightarrow 0}
\lim_{\partial_0\rightarrow 0}
S_{\rm eff}[a]
&=&
\int d^3x\left[\left(\nu\frac{e^2}{2\pi}+\alpha^{xy}+C_{l_0}\right)\frac{1}{2}
\epsilon^{\mu\nu\rho}a_\mu\partial_\nu a_\rho\right.\\
&&\left.+\frac{v_n^2}{2}\{(a^{\rm T}_0)^2-c_s^2(a_i^{\rm T})^2\}
\right].\nonumber
\end{eqnarray}
This limit is applicable to a physical situation when the gauge field is spatially inhomogeneous 
or dynamical. 
The corresponding infrared limit is given in Appendix \ref{example}. 

\subsection{Topological consideration}

\begin{figure}[hpt]
\includegraphics[width=9cm,clip]{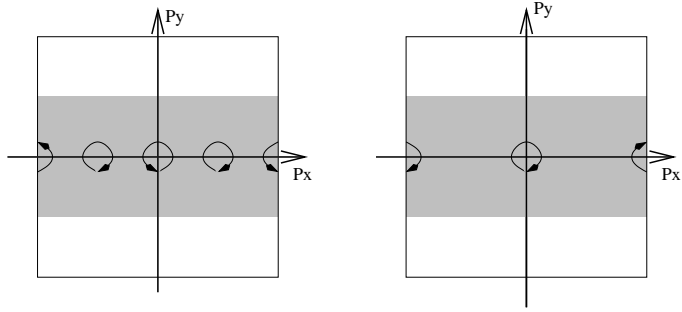}
\caption {Vortices of gap functions in the Fermi sea. The left figure is
 for $\varphi^{(3)}$ and the right one is for $\varphi^{(2)}$. 
The shaded portion is the Fermi sea of the striped Hall gas.
At the center
 of vortices, $\Delta(\ve{p})$ becomes zero. }
\label{fig:vortices}
\end{figure}

Finally, we study the topological nature of the correction term 
which depends on a global structure of a mean field solution. 
To see a topological property of the correction $\alpha^{ij}$ 
to the Chern-Simons coupling, 
let us consider the limit $\vert\Delta(\ve{p})\vert\rightarrow0$. 
In this limit, the quantity of the integral Eq.~(\ref{eq:alphaij}) 
comes from only the Fermi energy regime. 
Actually, we can show that
\begin{equation}
\lim_{\vert\Delta(\ve{p})\vert\rightarrow0}\int_{T^2}
d^2p \frac{\vert\Delta({\bf p})
\vert^2}{2\{g({\bf p})\}^3}f(\ve{p})
=\int_{\rm FS}\frac{dl}{(2\pi)^2}
\frac{f(\ve{p}(l))}{\vert\partial_{\bot}g_{z}\vert},
\label{eq:delta_gap}
\end{equation}
where $f(\ve{p})$ is an arbitrary function on the MBZ, 
the integral $\int_{\rm FS} dl$ is defined on the Fermi surface (line) $\ve{p}(l)$,  
and $\partial_{\bot}g_3$ is a Fermi velocity $v_F$. 
By substituting $\Delta(\ve{p})=\vert\Delta(\ve{p})\vert e^{i\Theta(\ve{p})}$ 
into $\alpha^{ij}$  
and using Eq.~(\ref{eq:delta_gap}), we obtain 
\begin{eqnarray}
\epsilon_{ij}\alpha^{ij}&=&\frac{e^2}{8\pi^2}\int_{\rm FS}
(\nabla\Theta(\ve{p})+\mathcal{A}(\ve{p}))\cdot d\ve{p}
\nonumber
\\
&=&\frac{e^2}{2\pi}\left(\frac{N_{v}}{2}+\nu_*\right).
\label{eq:topo_n}
\end{eqnarray}
Here the integer $N_v$ is a vortex number 
of $\Delta({\bf p})$ and 
$\nu_*$ is a filling factor in the $l_0$-th Landau level. 
In other words, the correction term $\epsilon_{ij}\alpha^{ij}$ is composed of 
the topological number $N_v$, and geometrical number $\nu_*$. 
If we use Eqs.~(\ref{eq:kai1}) and (\ref{eq:kai2}) 
where the Fermi surfaces are at 
$p_y=\pm\pi/2$, then one finds $N_v=0$, $\nu_*=1/2$. 
Thus we obtain $\alpha^{xy}=e^2/4\pi$ and $\alpha^{yx}=0$. 
If we use the $\varphi^{(2)}$ in place of $\varphi^{(3)}$ naively, 
then $N_v=-2$ 
and we obtain $\alpha^{xy}=-e^2/4\pi$ and $\alpha^{yx}=0$. 
The schematic figure of the vortex structures of $\Delta(\ve{p})$ is shown 
in Fig.~\ref{fig:vortices}. 
%%%%%%%%%%%%%%%%%%%%%%%%%%
%%%%%%%%%%%%%%%%%%%%%%%%%

In addition, in the limit $\vert\Delta(\ve{p})\vert\rightarrow0$, 
it is shown that 
\begin{eqnarray}
(c_s v_n)^2&=&0,\\
v_n^2&=&\mathcal{O}(1/v_F),\\
C_{l_0}&=&\mathcal{O}(\omega_c/v_F). 
\end{eqnarray}
In the striped Hall gas, the Fermi velocity logarithmically diverges due to the 
singularity of the Coulomb potential \cite{aoyama}. 
The non-universal quantities $v_n^2$ and $C_{l_0}$ are negligibly small in the limit. 
Therefore the correction term to the Chern-Simons coefficient in the limit 
$\vert\Delta(\ve{p})\vert\rightarrow0$ is given by the universal quantity in 
Eq.~(\ref{eq:topo_n}).

\section{Summary and Discussion}
\label{summary}

We have derived the low-energy effective action of a paired electron state in the 
quantum Hall system by using a field theoretical method. 
The derivation is performed in a gauge invariant way. 
In the one-loop order, the 
low-energy physics is described by only the Chern-Simons action whose
coupling constant is proportional to $\frac{e^2}{2\pi}\nu$. 
Beyond the one-loop order estimate, 
we have employed the WT identity whose infrared singular parts are accompanied
by the NG mode associated with the spontaneous $U(1)$ symmetry
breaking. 
Corrections are added to the coefficient of the Chern-Simons term according to 
an infrared limit. 
The mass term of the gauge field is induced. 
In a usual experiment observing the quantum Hall effect, 
the corresponding low-energy limit leads the correction term to be zero, 
which is expected in the Galilean invariant system \cite{Read_Green}. 
The correction term plays important roles when the gauge field 
is treated as a dynamical field. 
Spatially inhomogeneous static solutions of the gauge fields for vortex states, 
edge states, etc, lead the correction term to be non-zero. 

The Meissner term and corrections to the Chern-Simons coupling 
derived from the effective action in Eq.~(\ref{eq:Seffath}) 
could be observed by an experiment in the annular geometry (Corbino disk). 
By adiabatically inserting a half-flux quantum $\pi/e$ into the hole of the annulus, 
a supercurrent ${\bf j}_{\rm s}$ caused by the Meissner term circulates around 
the inner edge and an electric charge accumulates at the edge due to the 
Hall current induced by the electromagnetic induction \cite{LaughlinHalperin}. 
If there is no correction to the Chern-Simons coupling, then the accumulating charge is $
\pi/e\times\nu e^2/2\pi=e\nu/2$. 
Since the annulus is a multiply connected domain, the non-trivial winding mode $\theta_s$ 
can exist as a classical NG mode solution, 
which satisfies $\oint d\theta_{\rm s}=n\pi/e$ 
($n$: integer) where the integration path is in the bulk region. 
This winding mode saturates the supercurrent in the bulk region as ${\bf j}_{\rm s}=
(c_s v_n)^2(\BF{ a}+\nabla\theta_{\rm s})\approx0$ by the Meissner effect. 
The supercurrent flows at the edge region due to the spatial change of the 
order parameter. 
Since the order parameter fluctuates spatially at the edge region, 
the induced electromagnetic field is not uniform there. 
Then, the corrections to the Chern-Simons coupling become relevant and 
the accumulating charge deviates from $e\nu/2$ at the inner edge. 
Thus, observations of these phenomena would indicate the existence of the 
Meissner term and corrections to the Chern-Simons coupling obtained 
in this paper. 
To obtain a more quantitative prediction for these observations, we need to solve 
the dynamics of an electromagnetic field and the order parameter near the boundary 
of the annulus in the Ginzburg-Landau theory. 

Topological structure of the order parameter in various physical systems was 
investigated by Volovik \cite{Volovik}. 
The topological quantity of Eq.~(\ref{eq:topo_n}) reflects a 
global structure of the order parameter in the U(1) symmetry breaking 
quantum Hall system. 
This topological number may be observed  in some topological objects (ex. vortex). 
The charge and statistics of the vortex excitation are interesting future 
problems \cite{goryo2}.  
In the present case, the rotational invariance is broken and the vortex solution 
may become anisotropic. 
Furthermore, the existence of corrections to the Chern-Simons coupling beyond our
treatments, which comes from interactions between quasiparticles,
impurities, finite thickness, finite size \cite{finiteS}, finite temperature \cite{finiteT}, etc. 
is also still remained as an interesting future subject.

\begin{acknowledgements}
We thank J. Goryo for instructive discussions. 
T. A. is grateful for postdoctoral fellowship 
supported by ISSP. 
N. M. was partially supported by the special Grant-in-Aid for Promotion of 
Education and Science in Hokkaido University, 
a Grant-in-Aid for Scientific Research on Priority Area 
(Dynamics of Superstrings and Field Theories, Grant No. 13135201) and 
(Progress in Elementary Particle Physics of the 21st Century through Discoveries of 
Higgs Boson and Supersymmetry, Grant No. 16081201), provided by the 
Ministry of Education, Culture, Sports, Science, and Technology, Japan. 

\end{acknowledgements}

\appendix

\section{Notations and Useful relations}
\label{notation}

\subsection{Pauli's matrix}
The Pauli's matrix $\tau_i$ is defined by
\begin{equation}
\tau_1=\left(
\begin{array}{cc}
0&1\\
1&0 \\
\end{array}
\right),
\tau_2=\left(
\begin{array}{cc}
0&-i\\
i&0 \\
\end{array}
\right),
\tau_3=\left(
\begin{array}{cc}
1&0\\
0& -1 \\
\end{array}
\right).
\end{equation}
The trace formulae of these matrices are 
\begin{eqnarray}
&&
{\rm Tr}^{ph}[\tau_i]=0, {\rm Tr}^{ph}[\tau_i\tau_j]=2\delta_{ij}, 
{\rm Tr}^{ph}[\tau_i\tau_j\tau_k]=2i\epsilon_{ijk},
\nonumber
\\
&&
{\rm Tr}^{ph}[\tau_i\tau_j\tau_k\tau_l]=
2(\delta_{ij}\delta_{kl}-\delta_{ik}\delta_{jl}+\delta_{il}\delta_{jl}).
\label{eq:Pauli_tr}
\end{eqnarray}

\subsection{Proof of 
$e^{i \hat{\ve{k}}\cdot\ve{D}} f(p)=
e^{\frac{i}{2\pi}\hat{k}_x(\hat{k}_y+2p_y)}f(p+\hat{k})
$}
When $[X,Y]$ commutes with $X$ and $Y$, we obtain 
the Hausdorff formula
\begin{equation}
e^X e^Y=e^{[X,Y]}e^Y e^X,\ \ {\rm and }\  \ e^X
 e^Y=e^{\frac{1}{2}[X,Y]}e^{X+Y}.
\label{eq:Haus}
\end{equation}
By using this formulae, 
we can prove 
\begin{equation}
e^{i \hat{\ve{k}}\cdot\ve{D}} f(p)=
e^{\frac{i}{2\pi}\hat{k}_x(\hat{k}_y+2p_y)}f(p+\hat{k})
\label{eq:D}
\end{equation} as 
\begin{eqnarray*}
e^{i \hat{\ve{k}}\cdot\ve{D}} f(p)
=
e^{i \hat{\ve{k}}\cdot {\bf \mathcal{A} } } 
e^{i \hat{\ve{k}}\cdot\ve{p} }
e^{-\frac{1}{2}[i\hat{\ve{k}}\cdot{\bf \mathcal{A}}, 
i\hat{\ve{k}}\cdot\ve{p}]}
f(p)
=
e^{\frac{i}{2\pi}\hat{k}_x(\hat{k}_y+2p_y)}
e^{\hat{\ve{k}}\cdot\ve{p}} f(\ve{p})
=
e^{\frac{i}{2\pi}\hat{k}_x(\hat{k}_y+2p_y)}
f(p+\hat{k}),
\end{eqnarray*}
where we used 
$[i\hat{\ve{k}}\cdot{\bf \mathcal{A}}, i\hat{\ve{k}}\cdot\ve{p}]
=-i\frac{k_xk_y}{\pi}$.

\section{LL matrix elements}
\label{LLmatrix}
\subsection{Current vertex}
We present useful matrix elements between different LLs. 
The basic matrix elements are give by 
\begin{equation}
\langle l_1 |e^{i \BFS{q}\cdot\BFS{\chi}}| l_2
 \rangle\\
=
\left\{
\begin{array}{ll}
\sqrt{\frac{l_1!}{l_2!}}
\left(
\frac{i q_x-q_y}{\sqrt{4\pi}}
\right)^{l_2-l_1}e^{-\frac{q^2}{8\pi}}L_{l_1}^{l_2-l_1}
\left(
\frac{q^2}{4\pi}
\right)
& \hbox{for $l_2>l_1$}
\\
\sqrt{\frac{l_2!}{l_1!}}
\left(
\frac{iq_x+q_y}{\sqrt{4\pi}}
\right)^{l_1-l_2}e^{-\frac{q^2}{8\pi}}L_{l_2}^{l_1-l_2}
\left(
\frac{q^2}{4\pi}
\right)
& \hbox{for $l_1>l_2$}\\
e^{-\frac{q^2}{8\pi}}L_{l_1}
\left(
\frac{q^2}{4\pi}
\right)
& \hbox{for $l_2=l_1$}
\end{array}
\right.
.
\end{equation}
Then, 
$
f_{l_1,l_2}^{0}(q)=\langle l_1|e^{i \BFS{q}\cdot\BFS{\chi}}  |l_2 \rangle
$
, 
$
f_{l_1,l_2}^{x}(q)=i\omega_c \frac{\partial}{\partial q_y}
\langle l_1|e^{i \BFS{q}\cdot\BFS{ \chi}}  |l_2 \rangle
$
and
$
f_{l_1,l_2}^{y}(q)=-i\omega_c \frac{\partial}{\partial q_x}
\langle l_1|e^{i \BFS{q}\cdot\BFS{ \chi}}  |l_2 \rangle
$
are given by
\begin{eqnarray}
f_{l_1,l_2}^{0}(0)
&=&
\delta_{l_1,l_2},
\label{eq:f0}
\\
f_{l_1,l_2}^{x}(0)
&=&
\left.
i\omega_c \frac{\partial f_{l_1,l_2}^{0}(q)}{\partial q_y}\right|_{q=0}
=
\left\{
\begin{array}{ll}
-i \omega_c \sqrt{\frac{l_1+1}{4\pi}}
\delta_{l_2,l_1+1}& \hbox{for $l_2>l_1$}\\
i\omega_c\sqrt{\frac{l_1}{4\pi}}
\delta_{l_1,l_2+1}& \hbox{for $l_1>l_2$}\\
0 & \hbox{for $l_1=l_2$}
\end{array}
\right. ,
\label{eq:fx}
\\
f_{l_1,l_2}^{y}(0)
&=&
\left.
-i\omega_c \frac{\partial f_{l_1,l_2}^{0}(q)}{\partial q_x}\right|_{q=0}
=
\left\{
\begin{array}{ll}
\omega_c\sqrt{\frac{l_1+1}{4\pi}}
\delta_{l_2,l_1+1}& \hbox{for $l_2>l_1$}\\
\omega_c\sqrt{\frac{l_1}{4\pi}}
\delta_{l_1,l_2+1}& \hbox{for $l_1>l_2$}\\
0 & \hbox{for $l_1=l_2$}
\end{array}
\right..
\label{eq:fy}
\end{eqnarray}
The derivatives of the matrix element 
$\left.\partial_\nu f^{\mu}_{l_1,l_2}(q)\right\vert_{q=0}$ are obtained as 
\begin{eqnarray}
\left.
\frac{\partial f_{{l_1,l_2}}^{x}(q)}{\partial q^y}
\right|_{q=0}
&=&
\left\{
\begin{array}{ll}
i\omega_c
\frac{\sqrt{(l_1+1)(l_1+2)}}{4\pi}\delta_{l_2,l_1+2}
 &  \hbox{for $l_2>l_1$}\\
i\omega_c
\frac{\sqrt{l_1(l_1-1)}}{4\pi}\delta_{l_1,l_2+2}
 &  \hbox{for $l_1>l_2$}\\
-i\frac{E_{l_1}}{2\pi} \delta_{l_1,l_2} & \hbox{for $l_1=l_2$}\\
\end{array}
\right.
,
\label{eq:fxy}
\\
\left.
\frac{\partial f_{{l_1,l_2}}^{y}(q)}{\partial q^x}\right|_{q=0}
&=&
\left\{
\begin{array}{ll}
i\omega_c
\frac{\sqrt{(l_1+1)(l_1+2)}}{4\pi}\delta_{l_2,l_1+2}
  &  \hbox{for $l_2>l_1$}\\
i\omega_c
\frac{\sqrt{l_1(l_1-1)}}{4\pi}\delta_{l_1,l_2+2}
  &  \hbox{for $l_1>l_2$}\\
 i\frac{E_{l_1}}{2\pi} \delta_{l_1,l_2}
 & \hbox{for $l_1=l_2$}
\end{array}
\right.
.
\label{eq:fyx}
\end{eqnarray}
Immediately, 
the current vertex parts at the same momentum are given by 
\begin{eqnarray}
\Upsilon^{0}_{l_1l_2}(p,p)
&=&
\delta_{l_1l_2}\tau_3,
\\
\Upsilon^{x}_{l_1l_2}(p,p)
&=&
\left\{
\begin{array}{ll}
-i\omega_c\sqrt{\frac{l_2}{4\pi}}\delta_{l_2,l_1+1}\tau_0
& \hbox{for $l_1<l_2$}
\\
i\omega_c\sqrt{\frac{l_2+1}{4\pi}}\delta_{l_1,l_2+1}\tau_0
& \hbox{for $l_1>l_2$}
\end{array}
\right. ,
\\
\Upsilon^{y}_{l_1l_2}(p,p)
&=&
\left\{
\begin{array}{ll}
\omega_c\sqrt{\frac{l_2}{4\pi}}\delta_{l_2,l_1+1}\tau_3
& \hbox{for $l_1<l_2$}
\\
\omega_c\sqrt{\frac{l_2+1}{4\pi}}\delta_{l_1,l_2+1}\tau_3
& \hbox{for $l_1>l_2$}
\end{array}
\right..
\end{eqnarray}
The derivatives of the current vertices are also given by 
\begin{eqnarray}
& &
\partial^\rho \Upsilon^{\mu}_{l_1l_2}(p,p-\hat{q})|_{q=0}
-\partial^\rho \Upsilon^{\mu}_{l_1l_2}(p-\hat{q},p)|_{q=0}
\nonumber \\
&=&
2\partial^\rho \Upsilon^{\mu}_{l_1l_2}(p,p-\hat{q})|_{q=0}
\nonumber \\
&=&
\Big\{ 
(\partial^\rho f_{{l_1,l_2}}^{\mu}(q)|_{q=0}
-
\partial^\rho f_{{l_2,l_1}}^{\mu}(q)|_{q=0}
)
-
(
f_{{l_1,l_2}}^{\mu}(0)+f_{{l_2,l_1}}^{\mu}(0)
)
\frac{i\tilde{p}_y}{2\pi}\delta^{\rho x}
\Big\}\tau_0
\nonumber \\
& &
+
\Big\{ 
(\partial^\rho f_{{l_1,l_2}}^{\mu}(q)|_{q=0}
+
\partial^\rho f_{{l_2,l_1}}^{\mu}(q)|_{q=0}
)
-
(
f_{{l_1,l_2}}^{\mu}(0)-f_{{l_2,l_1}}^{\mu}(0)
)
\frac{i\tilde{p}_y}{2\pi}\delta^{\rho x}
\Big\}\tau_3.
\label{eq:deriU}
\end{eqnarray}

\subsection{Unitary matrix $U(\ve{p})$}

By using the relation $[\xi,\eta]=\frac{1}{2\pi i}$, we obtain
\begin{eqnarray}
U^\dagger({\bf p})\xi U({\bf p})&=&\xi-\frac{p_y}{2\pi},\\
U^\dagger({\bf p})\eta U({\bf p})&=&\eta+\frac{p_x}{2\pi},\\
U^\dagger({\bf p})\partial^x U({\bf p})&=&-i\xi+i\frac{p_y}{2\pi},\\
U^\dagger({\bf p})\partial^y U({\bf p})&=&-i\eta.
\end{eqnarray}

The unitary matrix in the Nambu representation satisfies
\begin{eqnarray}
&{\cal U}^\dagger({\bf p})\partial^x{\cal U}({\bf p})=
\left(
\begin{array}{cc}
-i\xi&0\\
0& -i\,^t\xi \\
\end{array}
\right)+i\frac{p_y}{2\pi}\tau_3,
\label{eq:uni_nambu1}
\\
&{\cal U}^\dagger({\bf p})\partial^y {\cal U}({\bf p})=
\left(
\begin{array}{cc}
-i\eta&0\\
0& -i\,^t\eta \\
\end{array}
\right),
\label{eq:uni_nambu2}
\\
&\left[{\cal U}^\dagger({\bf p})\partial^x{\cal U}({\bf p}),
{\cal U}^\dagger({\bf p})\partial^y {\cal U}({\bf
p})\right]=\frac{i}{2\pi}\tau_3.
\label{eq:uni_nambu3}
\end{eqnarray}

The unitary matrix and the Green's function satisfy
\begin{eqnarray}
[{\mathcal G}_{\rm reg},\tau_3]&=&0,\\
\left[{\cal U},\tau_3\right]&=&0,\\
{\cal U}^\dagger\partial^i({\cal U}{\mathcal G}^{-1}
{\cal U}^\dagger){\cal U}
&=&
\partial^i {\mathcal G}^{-1}+[{\cal U}^\dagger\partial^i{\cal U}, 
{\mathcal G}^{-1}].
\end{eqnarray}

\section{Examples of the Infrared Limit}
\label{example}

In this appendix, we give two kinds of infrared limit. 
First one is spatially homogeneous limit and second one 
is spatially inhomogeneous limit. 
Let us introduce a monotonically increasing function of time $f(t)$ 
which satisfies
\begin{equation}
f(\infty)=1,\ 
f'(\infty)=f''(\infty)=\cdots=0,\ 
f'(-\infty)=f''(-\infty)=\cdots=0.
\end{equation}
We consider an external electric field ${\cal E}_x=\partial_0 a_x-\partial_x a_0$, say,  
in $x$ direction.  
Let us suppose that 
the external field is absent at $t=-\infty$ and 
the uniform electric field ${\cal E}_x^{(\infty)}$ are applied at $t=\infty$. 
By using Eq.~(\ref{eq:eff_action}), the electric current $j_y$ is calculated by
\begin{equation}
ej_y=-\frac{\delta S_{\rm eff}}{\delta a_y}=\nu\frac{e^2}{2\pi}{\cal E}_x
-
\frac{c_s^2(-\epsilon^{ij}\alpha^{ki}\partial_k\partial_j+C_{l_0}\partial_i^2)}
{\partial_0^2-c_s^2\partial_i^2}
{\cal E}_x+(c_sv_n)^2 a_y.
\label{eq:jy}
\end{equation}
where a gauge fixing $\partial_0 a_0-c_s^2\partial_i a_i=0$ is used. 

\subsection{Spatially homogeneous infrared limit}

An example of the spatially homogeneous infrared limit is 
\begin{equation}
a_0=0,\ a_x={\cal E}_x^{(\infty)}\frac{1}{\partial_0}f(t),\ a_y=0,
\end{equation}
where $\frac{1}{\partial_0}f(t)=\int_{-\infty}^tdt'f(t')$. 
In this case,  ${\cal E}_x={\cal E}_x^{(\infty)}f(t)$ and the second term in Eq.~(\ref{eq:jy}) vanishes. 
Then, no correction appears in the Hall conductance. 

\subsection{Spatially inhomogeneous infrared limit}

An example of the spatially inhomogeneous infrared limit is 
\begin{equation}
a_0=-{\cal E}_x^{(\infty)} f(t)x,\ a_x=-\frac{{\cal E}_x^{(\infty)}}{c_s^2}f'(t)\frac{x^2}{2},\ a_y=0.
\end{equation}
In this case,   ${\cal E}_x=-\frac{{\cal E}_x^{(\infty)}}{c_s^2}f''(t)\frac{x^2}{2}+
{\cal E}_x^{(\infty)}f(t)$. 
Using the relation $\frac{1}{\partial_0^2}f''(t)=f(t)$, the second term  in Eq.~(\ref{eq:jy}) 
becomes $(\alpha^{xy}+C_{l_0}){\cal E}_x^{(\infty)}$ at $t=\infty$. 
Then, the correction appears in the Hall conductance. 

Thus, the space and time dependent electric field like the above one 
promises to detect the striking topological properties of 
gap functions.

\end{document}